\documentclass[epjST]{svjour}
\usepackage{graphicx}
\usepackage{amssymb}
\usepackage{color}

\begin{document}
\title{Prethermalization in one-dimensional Bose gases: description by a stochastic Ornstein-Uhlenbeck process}
\author{Tim Langen\inst{1}\and Michael Gring\inst{1}\and Maximilian Kuhnert\inst{1}\and Bernhard Rauer\inst{1}\and Remi Geiger\inst{1}\and David Adu Smith\inst{1}\and Igor E. Mazets\inst{1,2}\and J\"org Schmiedmayer\inst{1}\fnmsep\thanks{\email{schmiedmayer@atomchip.org}}}
\institute{Vienna Center For Quantum Science and Technology, Atominstitut, TU Wien, Stadionallee 2, 1020 Wien, Austria\and Ioffe Physico-Technical Institute of the Russian Academy of Sciences, 194021 St. Petersburg, Russia}
\abstract{
We experimentally study the relaxation dynamics of a coherently split one-dimensional Bose gas using matterwave interference. Measuring the full probability distributions of interference contrast reveals the prethermalization of the system to a non-thermal steady state. To describe the evolution of noise and correlations we develop a semiclassical effective description that allows us to model the dynamics as a stochastic Ornstein-Uhlenbeck process. } 

\maketitle

\section{Introduction}
\label{intro}

Non-equilibrium dynamics and relaxation processes in isolated quantum systems constitute a fundamental question in many fields of physics\,\cite{polkovnikov2011}. Open problems occur on vastly different energy, length and time scales, ranging from the dynamics in the early universe and heavy-ion collisions\,\cite{berges} to the subtle coherence and transport properties in condensed matter systems\,\cite{eckstein,moeckel,kollath,mathey,barnett,marino}. However, in contrast to thermal equilibrium situations both the theoretical understanding of non-equilibrium phenomena as well as the experimental techniques to study complex transient states are still in their infancy. In particular, it remains an open question how an isolated quantum many-body system reaches thermal equilibrium from a given initial non-equilibrium state, a question touching the very fundamentals of quantum statistical mechanics\,\cite{polkovnikov2011,deutsch,srednicki,rigol}. This relaxation has recently been shown experimentally to proceed via an intermediate non-thermal state\,\cite{gring} exhibiting prethermalization\,\cite{berges}, where the system already demonstrates certain features similar to thermal equilibrium, while it is still strongly non-equilibrium.  In\,\cite{gring}, this decay was modelled using a Tomonaga Luttinger liquid formalism\,\cite{kitagawa,kitagawa2011}. Here we show that the correlations in the system during the relaxation can be described by an effective model based on an Ornstein-Uhlenbeck stochastic process\,\cite{uhlenbeck}.

The system we study is a repulsive, one-dimensional (1D) Bose gas in the quasi-condensate regime\,\cite{petrov,esteve}. In this regime density fluctuations are suppressed, and noise and correlations are determined mainly by strong phase fluctuations along the longitudinal axis of the system. This is in contrast to true three-dimensional Bose-Einstein condensates where long-range order allows the characterization of the order parameter by a single phase. As we will see in the following these phase fluctuations can lead to complex dynamics in 1D Bose gases. 

\begin{figure}[tb]
	\centering
	\includegraphics[width=0.90\textwidth]{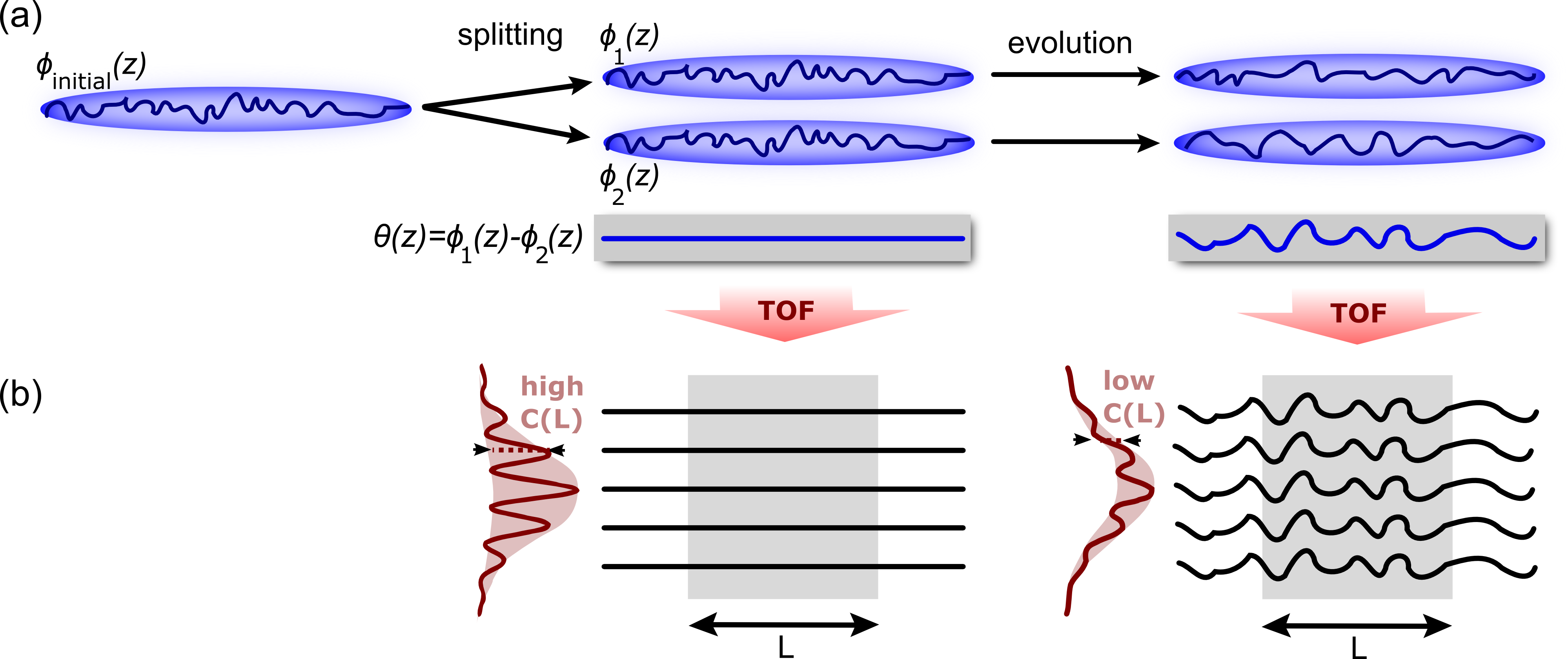}
	\caption{Non-equilibrium dynamics of a coherently split 1D Bose gas. An initial phase fluctuating 1D Bose gas is rapidly split into two uncoupled gases, such that each gas has almost the same phase profile $\phi_1(z)\simeq\phi_2(z)$. Right after splitting, the relative phase profile $\theta(z)$ is thus almost flat, reflecting the strong correlations between the gases. During the evolution of the system these correlations are expected to be lost, leading to a randomization of the relative phase profile, which can be directly studied using the corresponding decrease of the longitudinally integrated interference contrast $C(L)$ in time-of-flight.}
	\label{fig:1}
\end{figure}

To initialize the non-equilibrium dynamics, we rapidly split an initial single 1D Bose gas coherently into two uncoupled gases with almost identical longitudinal phase profiles\,\cite{kitagawa,kitagawa2011,bistritzer}. This situation is depicted in Fig.\,(\ref{fig:1}a). While the two resulting gases contain all the thermal entropy of the initial gas in their \textit{common} degrees of freedom (e.g. $\theta_\mathrm{com}(z) = \phi_1(z)+\phi_2(z)$), the \textit{relative} phase profile $\theta(z) = \phi_1(z)-\phi_2(z)$ is influenced only by very weak fluctuations resulting from the splitting process. Here, $\phi_{1,2}$ denote the longitudinal phase profiles of the first and second gas after splitting, respectively. As these common and relative degrees of freedom are populated according to completely different energies, the system is expected to relax, eventually leading to a randomization of the relative phase profile\,\cite{burkov,mazets2008,mazets2010,tan}. The question we address here is how far this randomization will proceed. In other words, will the two gases completely forget their initial correlations as commonly associated with the approach to thermal equilibrium, or not? 

In our experiment the evolution can directly be observed by performing matterwave interferometry in time-of-flight\,\cite{schumm,hofferberth2008,hofferberth2007}, where the integrated contrast of the interference pattern is directly related to the relative in-situ phase profile as shown in Fig.\,(\ref{fig:1}b). We will demonstrate in the following that this allows for a complete characterization of the transient states that are reached during the evolution. 

\section{Experiment}
\label{sec:1}
\begin{figure}[tb]
	\centering
	\includegraphics[width=1.0\textwidth]{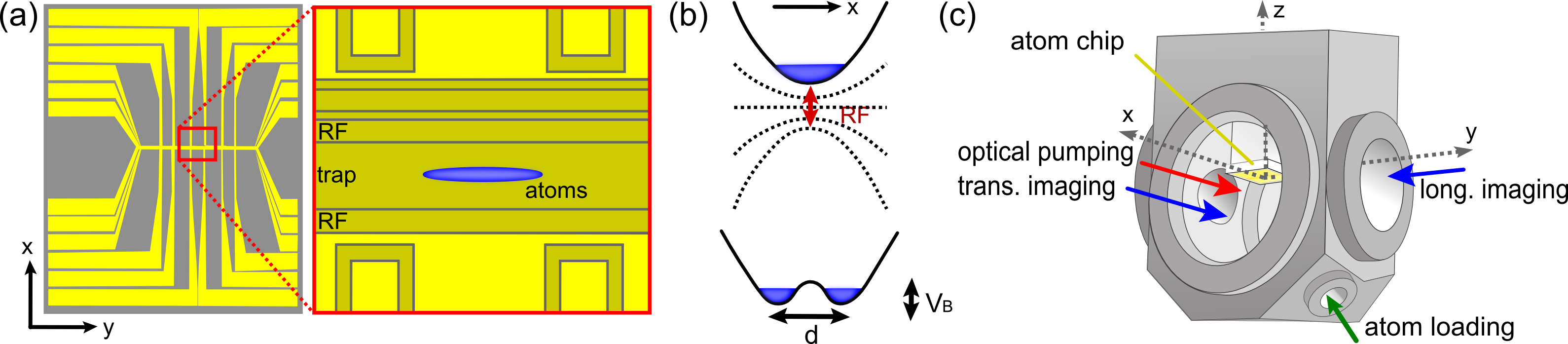}
	\caption{Experimental setup. (a) A single 1D Bose gas is realized on an atom chip. The atoms are initially trapped in a static magnetic trap formed about $100\,\mu$m below a micro-fabricated wire. Longitudinal confinement is provided by additional U-shaped wires. (b) The resultant harmonic trapping potential can be smoothly deformed into a double well trap with separation $d \simeq 3.5\,\mu$m and barrier height $V_B\simeq 3\,$kHz by coupling the atomic Zeeman levels of the $F=2$ manifold using radio-frequency radiation which is applied via parallel wires adjacent to the central trapping wire. This allows the realization of the initial non-equilibrium state described in Fig. (\ref{fig:1}). (c) Experimental chamber with the atom chip mounted upside down in the center. Two absorption imaging systems are used to probe the gas longitudinally (matterwave interference) and transversally (temperature measurement). The latter imaging system is also used for the optical pumping scheme described in Fig.(\ref{fig:3}).}
	\label{fig:2}
\end{figure}

To prepare a single 1D Bose gas, we use standard evaporative cooling in a magnetic microtrap on an atom chip\,\cite{reichel}. The experimental setup is depicted in Fig. (\ref{fig:2}). We typically trap degenerate gases containing $2$-$10 \times 10^3$ ${}^{87}$Rb atoms in the $F=2, m_F=2$ state. This corresponds to a peak linear density of $n_\mathrm{1D}\sim20-80\,\textrm{atoms}/\mu$m. The proximity to the current carrying structures of the atom chip allows the realization of very tight radial trapping such that the gas becomes very anisotropic and behaves quasi-one-dimensional\,\cite{esteve,krueger}, i.e. $k_B T,\mu<\hbar\omega$, where $\mu\sim 300-1000\,$Hz is the chemical potential of the gas and $\omega \simeq 2\pi\times2\,$kHz is the typical radial trapping frequency. The temperature $T\sim 20-150\,$nK is determined from measurements of the two-point density correlation function of the atoms obtained by imaging them transversally in time-of-flight\,\cite{imambekov,manz,smith}. 

\begin{figure}[tb]
	\centering
	\includegraphics[width=0.80\textwidth]{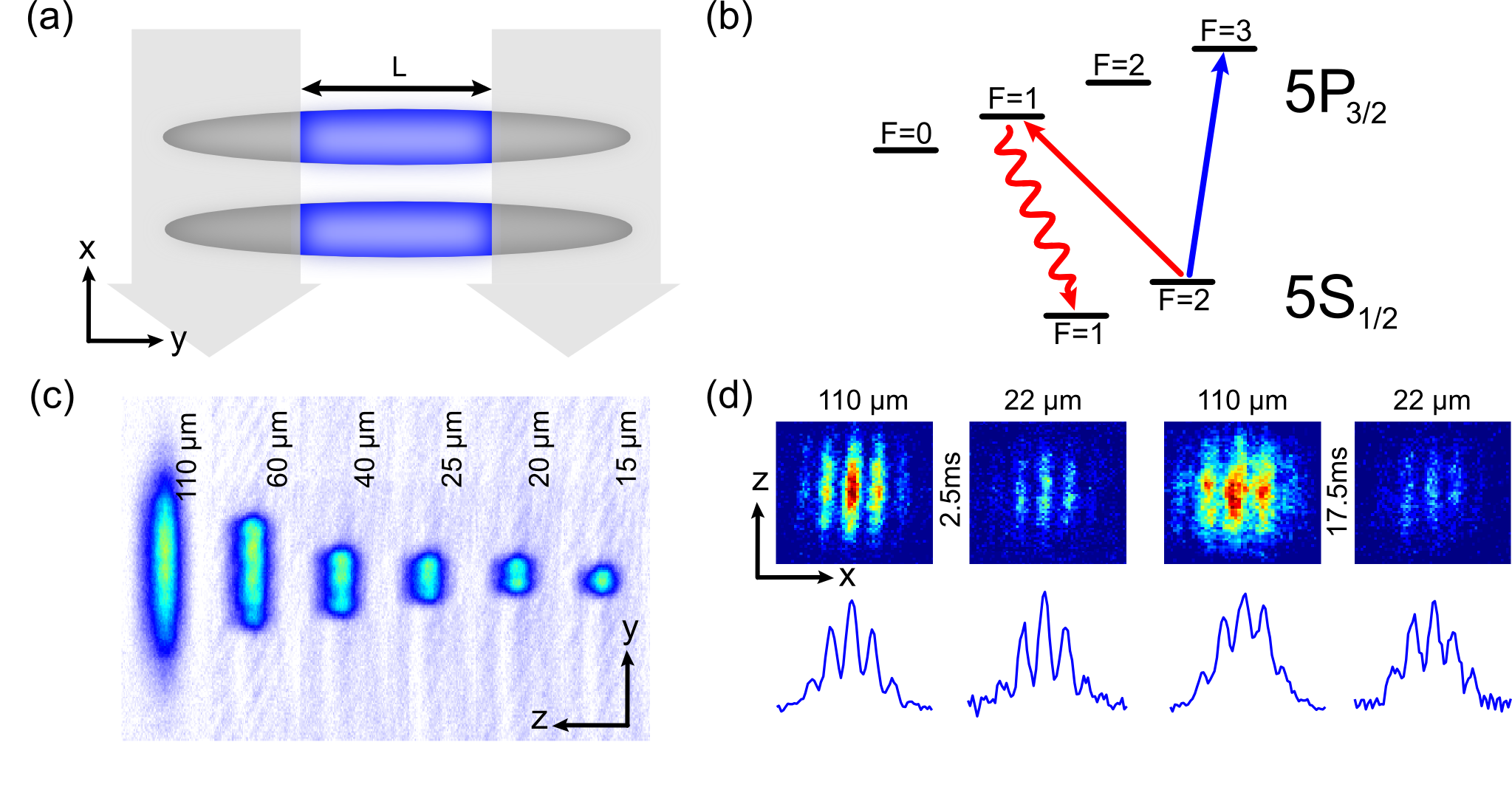}
	\caption{Spatially selective matterwave interference. (a) To change the integration length $L$ in the imaging process the cloud is transversally exposed to a $175\,\mu$s-long optical pumping pulse after $1\,$ms time-of-flight (TOF). This renders the edges of the cloud invisible to the imaging process. The desired length $L$ is protected from the optical pumping by imaging the shadow of a variable sized target onto the center of the cloud. (b) Atomic level scheme for imaging and optical pumping. Imaging is performed on the $F=2 \rightarrow F'=3$ transition of the $\textrm{D}_2$ line. For the optical pumping we employ the $F=2 \rightarrow F'=1$ transition, from where the atoms predominantly decay to the dark $F=1$ state after scattering only very few photons. (c) Result of the optical pumping for various integration lengths $L$ observed after $12\,$ms TOF expansion using the transversal absorption imaging system. (d) Typical longitudinal absorption pictures and corresponding line profiles for $L=110\,\mu$m and $L=22\,\mu$m after $2.5\,$ms and $17.5\,$ms of evolution. The contrast $C(L)$ is quantitatively extracted by fitting the line profiles with a periodically modulated Gaussian. The shortest $L$ possible is determined by the remaining number of visible atoms, which decreases with decreasing $L$. }
	\label{fig:3}
\end{figure}

To realize the non-equilibrium experiment introduced above, we deform the static harmonic trap of the atoms into a fully controllable double well potential by applying radio-frequency radiation via additional wires on the atom chip\,\cite{hofferberth2007,lesanovsky}. The typical timescale for the decoupling of the two gases during the fast splitting process is on the order of $0.5\,$ms. After the splitting, we hold the atoms for a variable evolution time in the double well potential. To study the resulting matterwave interference pattern between the two halves of the system, all trapping potentials are turned off and the two gases expand and overlap in time-of-flight. Information is extracted using absorption imaging along the longitudinal axis of the cloud\,\cite{smith}. This configuration automatically performs an integration of the interference pattern as depicted in Fig. (\ref{fig:1}b). To realize a certain integration length $L$ in this process, we use an optical pumping scheme (see Fig. \ref{fig:3}) to spatially select atoms within a certain $L$. This procedure is technically simple to implement, very stable and avoids the complications connected to direct imaging of the interference pattern by reflection from the atom chip as discussed in\,\cite{gring}.

To study the dynamics we extract the interference contrast $C(L)$ as a function of $L$ and time. Typical interference pictures demonstrating the decay of contrast with time are shown in Fig.(\ref{fig:3}d). This decay has been studied in detail in\,\cite{gring,hofferberth2007,kuhnert}, showing a rapid evolution over a timescale of approximately $10\,$ms, followed by the emergence of a steady state. This observation demonstrates that the relative phase does indeed partially randomize. However, it does not provide any information about the processes that are responsible for the decay or on the nature of the steady state.  

To extract more information about the dynamics we thus go beyond simple mean values and measure the full quantum mechanical probability distribution function (FDF) of the contrast $C(L)$\,\cite{gring,kitagawa,hofferberth2008,polkovnikov2006,gritsev}. While the mean contrast can be connected to the two-point phase correlation function 
of the gas\,\cite{petrov,whitlock,betz}, the FDFs contain also information about all higher even correlation functions and consequently allow the characterization of many-body quantum states with much more detail. Experimentally, this has successfully also been used before to study 1D gases in\,\cite{hofferberth2008} and out of\,\cite{gring} thermal equilibrium, as well as the dynamics of an unstable quantum pendulum\,\cite{gerving}. 

To extract the FDFs, we repeat the experiment many times with the same initial conditions and measure the different outcomes of the contrast. Note that this procedure is fundamentally different to the averaging performed in a 2D optical lattice\,\cite{trotzky}. There, typically many 1D gases are realized and probed in parallel which means that only ensemble averages are accessible. Due to the central limit theorem the statistics of those average values is typically approximately Gaussian, meaning that the information contained in the higher moments (i.e. the special shape of the FDFs) is not accessible in this way.

\begin{figure}[tb]
	\centering
	\includegraphics[width=0.95\textwidth]{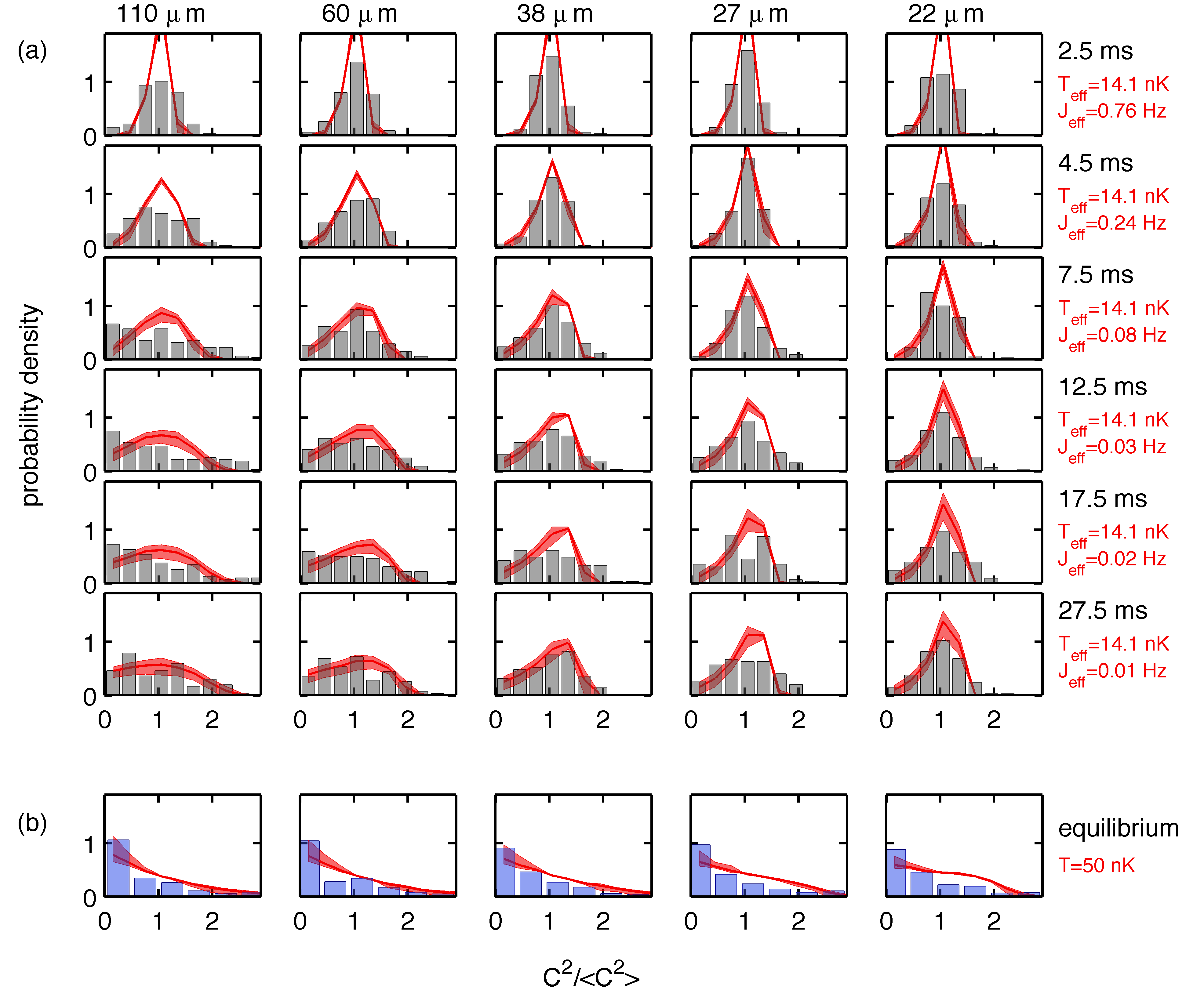}
	\caption{(a) Full distribution functions (FDFs) of contrast $C(L)^2$ (histograms) revealing the underlying physics behind the dynamics of the non-equilibrium system. We find very good agreement with the effective Ohrnstein-Uhlenbeck model described in the text (solid line).  Shaded areas denote the statistical uncertainty corresponding to the $170$ samples used in the experimental data.
The significantly different behavior on short and long length scales directly visualizes the multimode nature of 1D Bose gases\,\cite{kuhnert}.
	(b) Typical equilibrium FDFs measured by splitting a thermal gas and creating two completely independent quasi-condensates. The strong difference in shape of the thermal FDFs and the observed non-equilibrium steady state directly demonstrate that this steady state is not thermal equilibrium.}
		\label{fig:4}
\end{figure}

The results of our procedure are shown in Fig. (\ref{fig:4}). Initially we find peaked distributions on all length scales, directly reflecting the coherent initial state. The distributions develop markedly different forms for different $L$ during the evolution. For an integration over the whole length of the cloud ($L=110\,\mu$m) the distributions quickly develop an exponential shape. This exponential shape is characteristic for a state where the correlation length is much shorter than $L$. On long length scales, our interferometer thus seems to have lost all its initial correlations. On the other hand, on short length scales ($L=22\,\mu$m) the distributions remain peaked for all times probed allthough $L$ is still significantly larger than the thermal coherence length. This shows that some of the correlations of the initial state did indeed remain in the system. 

Similar to the mean contrast\,\cite{gring,kuhnert}, the FDFs change strongly during the first $10\,$ms of the evolution, but then reach a steady state. We can use the FDFs to directly show that the steady state the system reaches, is not thermal equilibrium. For a system of two uncoupled quasi-condensates the thermal equilibrium state is characterized by the fact that both gases are completely independent. One can show that in this case relative and common degrees of freedom have identical temperatures, as expected for a thermal equilibrium state. We can prepare this situation in experiment by splitting a thermal, non-condensed gas using the double well potential. We then perform evaporative cooling in the double well, producing two completely independent quasi-condensates which do not have any knowledge of each other\,\cite{hofferberth2008,betz}. Mapping the distributions we find the characteristic exponential decay on all length scales probed (see Fig.\,\ref{fig:4}b). This differs strongly from our observations for the steady-state of the non-equilibrium evolution of the coherently split system. We thus conclude, without the need to refer to any theoretical model that the observed state is not the true thermal equilibrium state of the system. 

\section{Understanding the dynamics}
\label{sec:2}
We now introduce a model to describe the evolution of the system and the nature of the steady state, showing that the system reaches a prethermalized state. Previously we have studied this system using a homogeneous, integrable Luttinger liquid description\,\cite{gring,kitagawa}, which has been shown to be a good approximation for the nearly-integrable dynamics of the trapped system. Here, we present a complementary approach based on an effective stochastic model using an Ornstein-Uhlenbeck process\,\cite{uhlenbeck,gardiner}, which provides valuable insights into the correlation properties of the gases, is numerically very efficient, and allows us to take into account technical properties of the experimental setup more easily. This approach has been previously introduced to study equilibrium properties of 1D\,\cite{stimming} and 2D\,\cite{mazets2012} gases, providing a straightforward link between equilibrium and non-equilibrium dynamics in our experiment.\\

A system of two uncoupled quasi-condensates can be described by the Hamiltonian
\begin{equation} 
\hat H=\int\,dz\left[\sum_{j=1}^2\left(\frac{\hbar^2}{2m}\frac{\partial \hat\Psi_j^\dag}{\partial z}\frac{\partial\hat\Psi_j}{\partial z}+\frac{g}{2}\hat\Psi_j^\dag\hat\Psi_j^\dag\hat\Psi_j\hat\Psi_j-\mu\,\hat\Psi_j^\dag\hat\Psi_j\right)\right],
\end{equation}  
where $g=2\hbar\omega a$ is the 1D interaction strength, $m$ is the atomic mass, $a\simeq5.23\,$nm is the 3D scattering length and $\mu=g n_\mathrm{1D}$ is the chemical potential. The index $j=1,2$ denotes the two gases and $n_\mathrm{1D}$ is the linear density in each one of them. We decompose the field operators into $\hat\Psi_{j}=\sqrt{\hat{n }_{j}(z)}\exp[{i\hat{\phi }_{j}(z)}]$. Here, $\hat \phi_{j} (z)$ and $\hat n_{j} (z)$ denote the operators describing the respective phase and linear density. This approach allows us to employ a generalization of standard Bogoliubov theory to describe the system, as density fluctuations in the two individual gases are suppressed and can be considered as small corrections~\cite{mora,giamarchi}. 

To describe the correlations between the two gases that are introduced in the splitting it is convenient to transfer the problem into symmetric and anti-symmetric coordinates which correspond to the common and relative degrees of freedom. We find that these degrees of freedom completely decouple in our system\,\cite{kitagawa2011}. Consequently, to describe the evolution of the interference contrast, we only have to study the relative degrees of freedom, where 
$\hat{\theta }(z)=\hat \phi_{1}(z)-\hat \phi_{2}(z)$
denotes the operator of the relative phase between the two quasi-condensates~\cite{bistritzer,polkovnikov}. The canonically conjugate variable is the relative number density 
$\hat{\nu }(z) =[\hat n_{1} (z)-\hat n_{2} (z) ]/2$. 
The expansion of $\hat{\theta }(z)$ and $\hat{\nu }(z)$ into plane waves reads
\begin{equation} 
\hat \theta (z) =\sum _{k\neq0} \hat \theta _k \frac {e^{ikz}}{\sqrt{\mathcal{L}}} , \quad 
\hat \nu (z) =\sum _{k\neq0} \hat \nu _k \frac {e^{ikz}}{\sqrt{\mathcal{L}}}.
\label{sc.3} 
\end{equation}  
where we have assumed periodic boundary conditions with $k=(2\pi/\mathcal{L})\,\times$~integer, $\mathcal{L}$ denoting the size of the system. Note that for the expansion coefficients $[\hat\theta_k^\dag,\hat\nu_{k'}]\equiv [\hat\theta_{-k},\hat\nu_{k'}]=-i\delta_{kk'}$. In these variables the linearized Hamiltonian for the relative degrees of freedom is equivalent to a system of independent harmonic oscillators with momenta $k$ and reads \,\cite{kitagawa2011,whitlock}
\begin{equation} 
\hat H=\sum_{k\neq0}\left[\frac{\hbar^2 k^2 n_\mathrm{1D}}{4m}\hat\theta_k^\dag\hat\theta_k+\left(g+\frac{\hbar^2k^2}{4 m n_\mathrm{1D}}\right)\hat\nu_k^\dag\hat\nu_k\right]
\label{hamiltonian}
\end{equation}  

In the next step, we characterize the initial state of the dynamics. In the experiment the splitting is performed fast in comparison to the timescale set by the inverse chemical potential $t_\mathrm{split}<\hbar/\mu=\xi _\mathrm{h}/c$, where $\xi _\mathrm{h} =\hbar /(mc)\sim 500\,$nm is the healing length and $c=\sqrt{g n_\mathrm{1D}/m}$ is the speed of sound.  Thus there is no time for the atoms to correlate beyond $\xi_\mathrm{h}$ along the length of the gases and the longitudinal density fluctuations are completely random. In particular, there is no correlation between modes with different momenta. Also, for each atom the decision of going to either one of the gases is random and uncorrelated, leading to a binomial distribution of atom number fluctuations in each small part of the 1D system and the respective minimum-uncertainty relative phase distribution. Initially, at $t=0$, the relative phase and density fluctuations are thus governed completely by the 
atomic shot noise and hence are given by $\langle \hat\nu_k^\dag \hat\nu_k \rangle\vert _{t=0} = n_\mathrm{1D}/2$ and $\langle \hat\theta_k^\dag \hat\theta_k \rangle\vert _{t=0} = 1/2n_\mathrm{1D}$. In contrast, the fluctuations of the ground state of the Hamiltonian (Eq. ~\ref{hamiltonian}) are given by $\langle \hat\nu_k^\dag \hat\nu_k \rangle\vert_\mathrm{GS} = n_\mathrm{1D}{\cal S}_k/2$ and $\langle \hat\theta_k^\dag \hat\theta_k \rangle\vert_\mathrm{GS} = 1/2n_\mathrm{1D}{\cal S}_k$, where ${\cal S}_k =(1+k^{2}\xi _\mathrm{h}^{-2})^{-1/2}$ is the static structure factor.
In other words, at $t=0$ we have a squeezed state: the relative phase fluctuations are suppressed with respect to their ground-state value. On the contrary, there is a huge excess noise of the number imbalance. 

In the subsequent evolution, this noise behaves semiclassically. In what follows, we substitute the operators $\hat\nu_k$ and $\hat\theta_k$ with classical fields $\nu_k$ and $\theta_k$, respectively. The initial conditions for these classical fields are given by $\langle |\theta _k |^2 \rangle \vert _{t=0}=0$ and $\langle |\nu _k |^2\rangle \vert _{t=0} =n_\mathrm{1D} (1-{\cal S}_k ) /2.$ To obtain analytic results, we approximate the latter expression as
\begin{equation} 
\langle |\nu _k |^2\rangle \vert _{t=0} =\left \{ 
\begin{array}{ll}  
n_\mathrm{1D}/2 , & |k| \leq \xi _\mathrm{h}^{-1} \\ 
0               , & |k| >    \xi _\mathrm{h}^{-1}
\end{array} 
\right. . 
\label{sc.12} 
\end{equation} 
The Heisenberg equations of motion following from Eq.~(\ref{hamiltonian}) in the semiclassical limit are
\begin{equation} 
\dot{\nu}_k =n_\mathrm{1D}\left( \frac { \hbar k^2}{2m}\right)  \theta _k\quad \textrm{,}\quad\dot{\theta }_k=-\frac 1{\hbar n_\mathrm{1D} }\left( \frac {\hbar ^2 k^2}{2m}+2mc^2\right) \nu _k . 
\label{sc.14} 
\end{equation} 
The solution yields 
\begin{equation} 
\theta _k(t) =-\frac { {\hbar ^2 k^2}/(2m)+2mc^2 }{n_\mathrm{1D} \hbar \omega _k }\nu _k(0)  
\sin \omega _k t , 
\label{sc.16} 
\end{equation} 
where the eigenfrequency of a mode with momentum $k$ is given by 
\begin{equation} 
\omega _k = \sqrt{\left( \frac { \hbar k^2}{2m}\right) \left( \frac { \hbar k^2}{2m}+\frac {mc^2}\hbar \right) } = c k\sqrt{1+\xi _\mathrm{h}^2 k^2/4} . 
\label{sc.17} 
\end{equation} 
Over time, the initially high density fluctuations periodically turn into phase fluctuations and dephase, which leads to the observed decrease in interference contrast in the course of the evolution. To describe the decay we calculate the two-point phase correlation function 
\begin{equation} 
{\cal C}(z- z^\prime )= \frac{\langle\hat\Psi_1(z) \hat\Psi^\dagger_2(z) \hat\Psi^\dagger_1(z^\prime) \hat\Psi_2(z^\prime)\rangle}{\langle|\hat\Psi_1(z)|^2\rangle\langle|\hat\Psi_2(z^\prime)|^2\rangle} \simeq \langle e^{i \theta (z)-i\theta (z^\prime ) }\rangle = 
e^{-\frac 12\langle [\theta (z)-\theta (z^\prime )]^2 \rangle }.
\label{sc.19} 
\end{equation} 
Writing $t$ explicitly as an argument of ${\cal C}$, Eq. (\ref{sc.19}) can be expressed as 
\begin{equation} 
{\cal C}(\bar z,t) = \exp \left[ -\int _0^\infty \frac {dk}{\pi }\, \langle |\theta _k(t)|^2 \rangle 
(1-\cos k \bar z) \right] , 
\label{sc.21}
\end{equation} 
where $\bar z = z-z'$ and $\theta _k(t)$ is given by Eq. (\ref{sc.16}). Recalling that significant semiclassical noise is present only in 
modes with $|k|\lesssim \xi _\mathrm{h}^{-1}$, we use the long-wavelength approximation for the mode eigenfrequencies: $\omega _k\approx ck$. Then 
\begin{eqnarray} 
{\cal C}(\bar z,t) &=& \exp \Bigg [ -2\int _0^{\xi _\mathrm{h}^{-1}} dk\, 
\frac {\sin ^2 (c k\, t)}{\pi n_\mathrm{1D}\xi _\mathrm{h}^2 k^2} \nonumber (1-\cos k \bar z) \Bigg ]\\
&=&\exp \Bigg \{ - \frac {2mc^2t}{\hbar {\cal K}} \Bigg [ \Xi (2c\xi _\mathrm{h}^{-1}t)+\frac {\bar z}{2ct}\Xi (\xi _\mathrm{h}^{-1}\bar z)-
\nonumber \\ && 
\frac 12 \left| 1-\frac {\bar z}{2ct}\right| \Xi ( |2ct -\bar z|\xi _\mathrm{h}^{-1} ) -
\frac 12 \left| 1+\frac {\bar z}{2ct}\right| \Xi ( |2ct +\bar z|\xi _\mathrm{h}^{-1} )\Bigg ] \Bigg \},
\label{sc.27} 
\end{eqnarray}
where ${\cal K} =\pi n_\mathrm{1D}\xi _\mathrm{h} $ is the Luttinger liquid parameter of the system, $\Xi(x) =(\cos x-1)/x+\mathrm{Si}\, x$ and $\mathrm{Si}\, x =\int _0^x dy\, y^{-1} \sin y $ is the sine integral.

For the sake of definiteness, we set $\bar z>0$ from now on. If we are not interested in details of the correlation function behavior on 
the experimentally unresolved length scale $\lesssim \xi _\mathrm{h}$, we can simultaneously assume $\xi _\mathrm{h}^{-1}ct\gg 1$, 
$\bar z\xi _\mathrm{h}^{-1} \gg 1$, and $|2ct-\bar z|\xi _\mathrm{h}^{-1}\gg 1$. In this limit we obtain 
\begin{equation} 
{\cal C}(\bar z,t) =\left \{ 
\begin{array}{lll} 
\exp [ - {\pi \bar z}/({2 {\cal K} \xi _\mathrm{h} }) ] , &      0\leq &\bar z<2ct   \\
\exp [ - {\pi mc^2t}/({\hbar {\cal K}}) ] ,     &      &\bar z\geq 2ct 

\end{array} 
\right.  \, . 
\label{sc.28} 
\end{equation} 
We see from Eq. (\ref{sc.28}) that the phase correlations first decrease exponentially and then stabilize at a certain 
level. The crossover between exponential decrease and constancy of ${\cal C}(\bar z,t)$ and the value of ${\cal C}(\infty ,t)$ are 
time-dependent. \\

The essence of our approach is now to to compare Eq. (\ref{sc.28}) to its counterpart ${\cal C_\mathrm{eq}}(\bar z)$ in the stationary regime of two tunnel-coupled quasi-condensates at finite temperature $T$\,\cite{whitlock}
\begin{equation} 
{\cal C_\mathrm{eq}}({\bar z})=\exp \left[ -\kappa _T l_J \left( 1- e^{-\bar z/l_J} \right) \right] , 
\label{sc.29} 
\end{equation} 
where $\lambda _T = \hbar^2n_\mathrm{1D}/mk_\mathrm{B}T=1/\kappa_T$ is the thermal coherence length at temperature $T$ characterizing the thermal fluctuations which lead to a random phase along the longitudinal axis. $l_J$ is the corresponding length scale of a coupling $J$ which locks the relative phase of the two quasi-condensates\,\cite{betz}. This comparison leads to a conclusion that the time-dependent correlation function ${\cal C}(\bar z,t)$ during the dephasing emulates that of a coupled equilibrium system with the effective temperature and effective time-dependent phase-locking length 
\begin{equation} 
T^\mathrm{eff} =\frac {mc^2}{2k_\mathrm{B}} \quad , \quad l^\mathrm{eff}_J(t)=2ct. 
\label{sc.30} 
\end{equation} 
A quasi-steady state is reached on a timescale $t_e\gg \tau $, where the typical evolution time is\,\cite{bistritzer} 
\begin{equation} 
\tau = {\cal K}\xi _\mathrm{h}/\pi c=\hbar {\cal K}/\pi m c^2= n_\mathrm{1D}\xi _\mathrm{h}^2/c.
\label{sc.32}
\end{equation} 
For the parameters of the dataset shown in Fig.(\ref{fig:4}) we find
$\tau\sim 5\,$ms, in good agreement with our experimental observations that the steady state is reached for $t_e\gtrsim 10\,$ms. The effective length scale is directly related to an effective coupling  
$J_\mathrm{eff}(t)=\hbar/4m\,l_J^\mathrm{eff}(t)^2$. In the limit $t_e\gg \tau $ the effective coupling vanishes and we recover exactly the prethermalized state predicted in\,\cite{kitagawa} and observed in\,\cite{gring}. Although the system is still strongly non-equilibrium, it appears thermal-like and can be characterized by an effective temperature $T_\mathrm{eff}$, which is independent of the initial temperature of the gas. In particular, common and relative degrees of freedom are indeed decoupled and have not yet equilibrated. All dynamics can be attributed to dephasing within the relative degrees of freedom\,\cite{gring}. In the approach to the prethermalized state, $l_J^\mathrm{eff}$ plays the role of a characteristic length scale over which the system forgets the initial correlations for a given evolution time. The scaling of $l_J^\mathrm{eff}$ is that of a light cone\,\cite{mathey,cheneau}, reflecting that correlations in this many-body quantum system spread with a finite velocity given by the speed of sound. The typical correlation length reached in the prethermalized state is
\begin{equation} 
\lambda_{\mathrm{eff}} = 2 c\tau = 2 {\cal K}\xi _\mathrm{h}/\pi  \equiv \lambda_T(T_\mathrm{eff}).
\label{sc.33} 
\end{equation} 
For our typical parameters $\lambda_\mathrm{eff}\gg\lambda_{T}$, showing again that the prethermalized state is characterized by a strong memory of the initial state, leading to much higher correlations than expected in thermal equilibrium\,\cite{kuhnert}.

Our effective model along with the assumption of Gaussian fluctuations enables us to simulate the FDFs for any evolution time $t_e$ and any sampling length greater then the healing length using an Ornstein-Uhlenbeck stochastic process~\cite{gardiner} that develops in space, along the major axis of the trap,
\begin{equation}
	\frac{d}{dz}\phi(z)=-l_J^{-1}\phi(z)+f(z).
\end{equation}
Here, $f(z)$ is a random diffusion force characterized by $\langle f(z)\rangle=0$ and 
$\langle f(z_1)f(z_2)\rangle=2\kappa_T\delta(z_1-z_2)$, $l_J^{-1}$ describes friction. In our case, the ratio $T_\mathrm{eff}/n_\mathrm{1D}$ of the effective temperature to the linear atomic density defines the diffusion and the effective inter-well coupling provides a restoring force. The particular choice of correlation properties of $f(z)$ is required to reproduce the correlation function given by Eq.~(\ref{sc.29})\,\cite{gardiner,stimming} and is in very good agreement with experiments\,\cite{betz}. The correlation function may differ from Eq.~(\ref{sc.29}) on extremely short length scales $<\xi_h$ and hence $f(z)$ deviates from white noise on this scale. However, this length scale is not optically resolvable in experiment and can therefore be neglected in the analysis. 

To construct the FDFs, we use an exact updating formula\,\cite{gillespie} to simulate individual phase profiles. Properties of the experimental setup such as finite imaging resolution, trapping potential and time-of-flight expansion are directly taken into account in this procedure\,\cite{betz}. A statistical analysis yields the FDFs. The method is found to be computationally fast and efficient, especially for finite-size atomic clouds. 

In Fig.\,(\ref{fig:4}) we compare the results of the Ornstein-Uhlenbeck simulation to our experimental data.
We find very good agreement for all length scales and evolution times probed, both for the non-equilibrium and equilibrium cases.

\begin{figure}[tb]
	\centering
	\includegraphics[width=0.700\textwidth]{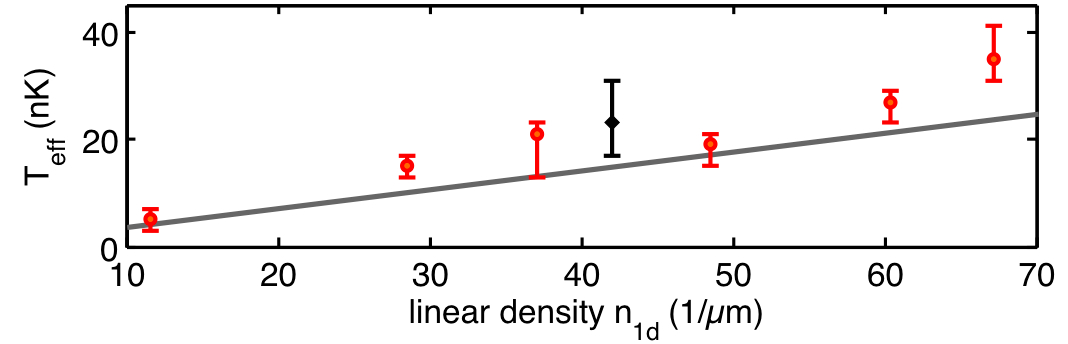}
	\caption{Scaling properties of the effective temperature. The fast splitting process leads to a linear dependence of $T_\mathrm{eff}$ on the density in each of the two gases, as given by 
	Eq. (\ref{sc.30}). The datapoint at $n_\mathrm{1D} = 41\,\mu$m${}^{-1}$ corresponds to the dataset presented in Fig (\ref{fig:4}). The remaining datapoints have been adapted from\,\cite{gring}.}
	\label{fig:5}
\end{figure}

All parameters have been independently extracted from experiment. Our approach directly reveals that the prethermalized state already shows thermal-like properties, while it is still strongly non-equilibrium. As predicted by our effective theory it can be described by an equilibrium distribution with an effective temperature. This temperature is determined only by the linear density and can be significantly lower than the true thermodynamical temperature describing the common degrees of freedom. In Fig.\,(\ref{fig:5}) we check the dependence of the effective temperature on the linear density of the gas, by fitting the experimentally obtained distributions with our model, leaving temperature as a free parameter. The results directly confirm the predicted linear scaling of Eq. (\ref{sc.30}), providing further evidence for our interpretation of the dynamics. 

\section{Conclusion}
In conclusion, we have demonstrated prethermalization of a coherently split 1D Bose gas. While the near-integrability strongly suppresses true thermalization of the system, dephasing within the relative degrees of freedom leads to the establishment of an intermediate steady-state which has thermal-like features but is notably different from true thermal equilibrium. These dephasing dynamics can be well approximated by a semiclassical effective model based on an Ornstein-Uhlenbeck stochastic process. 
Our theoretical approach can straight forwardly be extended to the study of coupled gases, which could be used to reveal the dynamical properties of the Sine-Gordon model. 
Recent experimental advances now enable us to also image the full spatial structure of the interference pattern\,\cite{gring,kuhnert}, enabling, for example, the direct study of two-point phase correlations\,\cite{betz}. In the future, we plan to exploit this enhanced toolbox to shed light on outstanding questions about the second timescale of the evolution, where integrability is expected to be broken and thermal equilibrium is finally established. 
	
\begin{acknowledgement}
This work was supported by the Austrian FWF through M1040-N16, the Doctoral Programme CoQuS (W1210) and the Wittgenstein Prize, the EU through the integrating project AQUTE, Siemens Austria, and The City of Vienna. I.E.M. acknowledges the financial support from the FWF (project P22590-N16). R.G. is supported by the FWF through the Lise Meitner-Programme (project P1423-N27).
\end{acknowledgement}

\end{document}